\documentstyle[twocolumn,aps,epsfig]{revtex}
\def\lesssim{\mathrel{\hbox{\rlap{\hbox{\lower4pt\hbox{$\sim$}}}\hbox{$<$}
}}}
\def\gtrsim{\mathrel{\hbox{\rlap{\hbox{\lower4pt\hbox{$\sim$}}}\hbox{$>$}}
}}

\begin{document}
\def\cmpl{\timeofday\ \today}

\title{Transposition of a local-density-dependent pion-nucleus potential to 
an effective density-linear potential -  generalized Seki-Masutani 
relations}

\author{Toshimitsu Yamazaki$^1$\footnote{E-mail: 
yamazaki@nucl.phys.s.u-tokyo.ac.jp}  and Satoru Hirenzaki$^2$}

\address{$^1$RI Beam Science Laboratory, RIKEN, 2-1 Hirosawa,
Wako-shi, Saitama, 351-0198 Japan}
\address{$^2$Department of Physics, Nara Women's University,
Kita-Uoya Nishimachi, Nara 630-8506, Japan}

\date{October 29, 2002}
\maketitle

\begin{abstract}
We have shown that a local-density-dependent term, $F[\rho(r)] \,\rho(r)$, 
of the pion-nucleus potential with a nuclear density, $\rho(r)$, can be 
transposed into a conventional density-linear term,  $F(\rho_e) \,\rho(r)$, 
with an effective nuclear density, $\rho_e$, which is found to be  close to 
$\sim 0.6\, \rho(0)$ for most $\pi^-$ bound states. The presently found 
relations, {\it generalized Seki-Masutani relations}, for the 
density-quadratic term,  the medium-modified isovector term and the 
double-scattering isoscalar term of the s-wave pion-nucleus interaction 
assure that the constant parameters in the conventional pion-nucleus 
potential are interpreted as being effective ones in the light of 
density-dependent effects.
\end{abstract}

\vspace{1cm}

\noindent
{\bf 1. Introduction}\\

The so called ``anomalous s-wave repulsion" in the pion-nucleus interaction 
has been a long-standing problem in pion physics (see, for instance, 
\cite{Batty97}). Many authors (see, for instance, \cite{Batty97,Konijn90}) 
have claimed the existence of a large extra repulsion arising from the 
density quadratic term, Re$B_0 \,\rho (r)^2$,  from global fits of pionic 
atom data to the  Ericson-Ericson potential \cite{Ericson66,Ericson}:
\begin{eqnarray}
V(r) &=& U_s (r) + U_p (r),\\ \nonumber
U_s (r) &=& -(2 \pi/m_{\pi}) \\ \nonumber
   & \times & [\epsilon_1 (b_0 \rho(r) + b_1 \Delta \rho(r)) +\epsilon_2 
B_0 \rho(r)^2], \\
\rho (r) &=& \rho_p (r) + \rho_n (r),\\
\Delta \rho (r) &=& \rho_n (r) - \rho_p (r),
\end{eqnarray}
where $U_s (r)$ and  $U_p (r)$ are for s-wave and p-wave parts, 
respectively, and $\rho_p (r)$ and $\rho_n (r)$  are the proton and neutron 
density distributions, respectively, and $\epsilon_1 = 1+ m_{\pi}/M_N$ and 
$\epsilon_2 = 1 +  m_{\pi}/2 M_N$ (hereafter, we adopt units of
  $m_{\pi}^{-1}$,  $m_{\pi}^{-1}$ and $m_{\pi}^{-4}$ for $b_0$, $b_1$ and 
Re$B_0$, respectively).
New light has been shed on this problem, namely, the possibility of an 
anomalous repulsion in the isovector part ($b_1$) as a unique signature of 
chiral symmetry restoration was pointed out \cite{Weise:00,Kienle:01}. In 
this respect, the deeply bound $\pi^-$ states which were produced in heavy 
nuclei (in $^{207}$Pb \cite{Yamazaki96,Yamazaki98,Gilg00,Itahashi00} and in 
$^{205}$Pb \cite{Geissel:01,Geissel:02a}) are of particular importance, 
since the binding energies and widths of 1s $\pi^-$  states depend nearly 
entirely on the s-wave potential ($U_s$) and, thus, only the 1s states in 
heavy nuclei ($N>Z$) provide unique information on the isovector part of 
the s-wave potential.  In fact, the above conjecture has been proved by 
using new information from the 1s $\pi^-$ state in $^{205}$Pb. The 
experiment indicates an enhanced isovector parameter, which is related to a 
decreased order parameter of chiral symmetry breaking in the nuclear 
medium, $b_1^{\rm free}/b_1 = {f_{\pi}^*}^2/{f_{\pi}}^2 = 
0.78^{+0.13}_{-0.09}$  \cite{Geissel:02a}. Recent global fits of pionic 
atom data  by Friedman \cite{Friedman:02a,Friedman:02b} have shown 
consistency with the above view.

Since the potential parameter $b_1$ is nuclear-density dependent in the 
theoretical context of Weise \cite{Weise:00}, a question can be  raised as 
to the mutual relation in adopting the conventional Ericson-Ericson 
potential with constant coefficients (model C) and  a modified 
local-density-dependent (model LDD) potential. In the present note we show 
that both methods are equally valid and are connected to each other, since 
a density-dependent term can be effectively linearized. Our procedure gives 
an answer to the following question: where do the bound pions probe the 
nuclear potential? This problem is closely related to the so-called {\it 
Seki-Masutani (SM) relation} \cite{SM}, which emphasized  a strong 
correlation between the density-linear isoscalar parameter ($b_0$) and the 
density-quadratic parameter (Re$B_0$). In a sense, we generalize the SM 
relation. \\
 
\noindent
{\bf 2. Where does a bound $\pi^-$ probe the nuclear potential?}\\

First, we examine which part ($r$) of the nuclear density, $\rho (r)$, is 
probed by a bound $\pi^-$.  Two typical nuclei, $^{16}$O and $^{208}$Pb, 
are considered. The former possesses only shallow bound states, whereas the 
latter accommodates deeply bound states.  The nuclear matter densities are 
assumed to take a 2-parameter Fermi distribution with a half-density radius 
($c_p$ and $c_n$) and a diffuseness parameter ($z_p$ and $z_n$). The 
parameter values were chosen as follows:
$c_p=6.652$ fm, $c_n=6.892$ fm, and $a_p=a_n=0.5234$ fm for $^{208}$Pb, and 
$c_p=c_n=2.482$ fm and $a_p=a_n=0.5234$ fm for $^{16}$O. The central 
nucleon densities are: $\rho_0$ = 0.150 for $^{208}$Pb and 0.172 for 
$^{16}$O. We take known potential parameters for the p-wave parts 
\cite{Batty97} and a set of s-wave parameters ($b_0=-0.028$, $b_1 = -0.12$, 
Re$B_0 =0$ and Im$B_0 = 0.055$).

\begin{figure}[hbtp]
\begin{center}
\vspace{0cm}
\epsfig{file=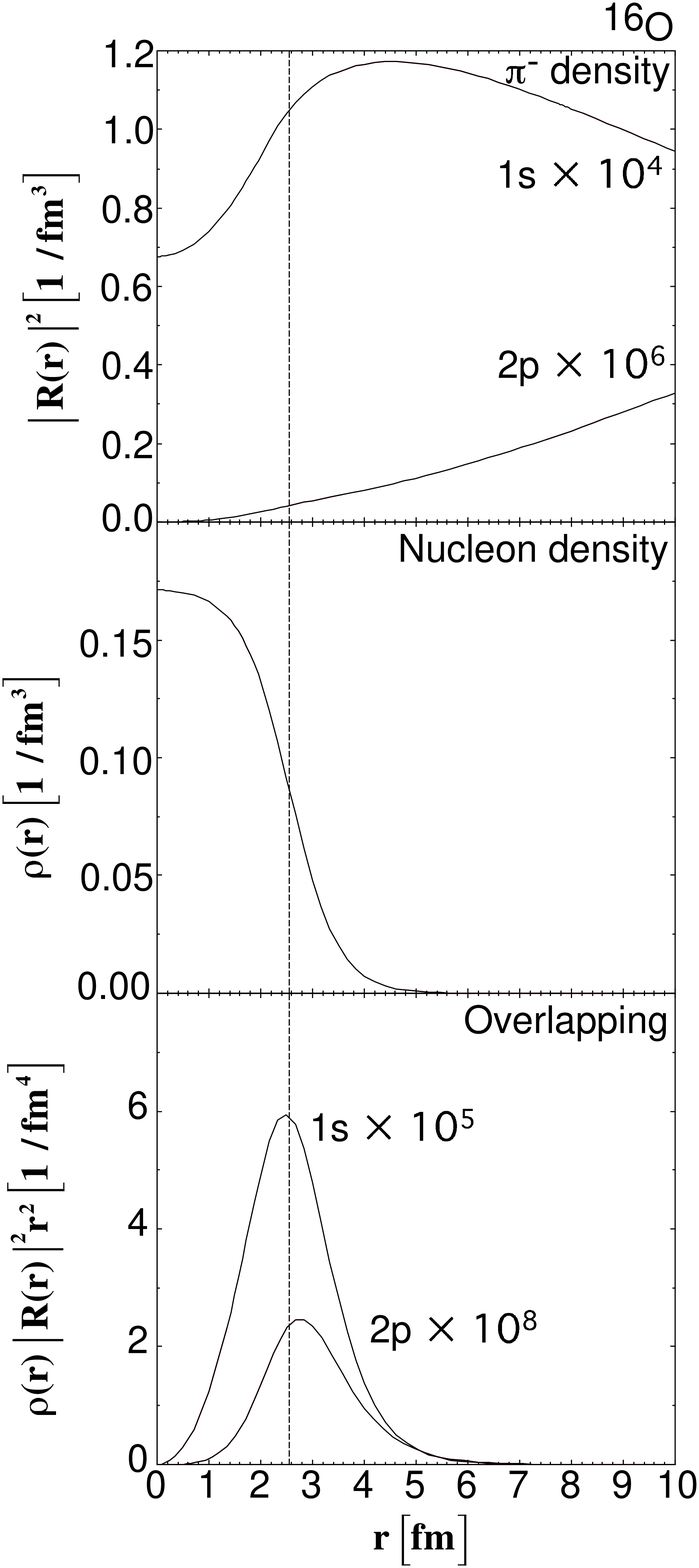,height=14cm}
\epsfig{file=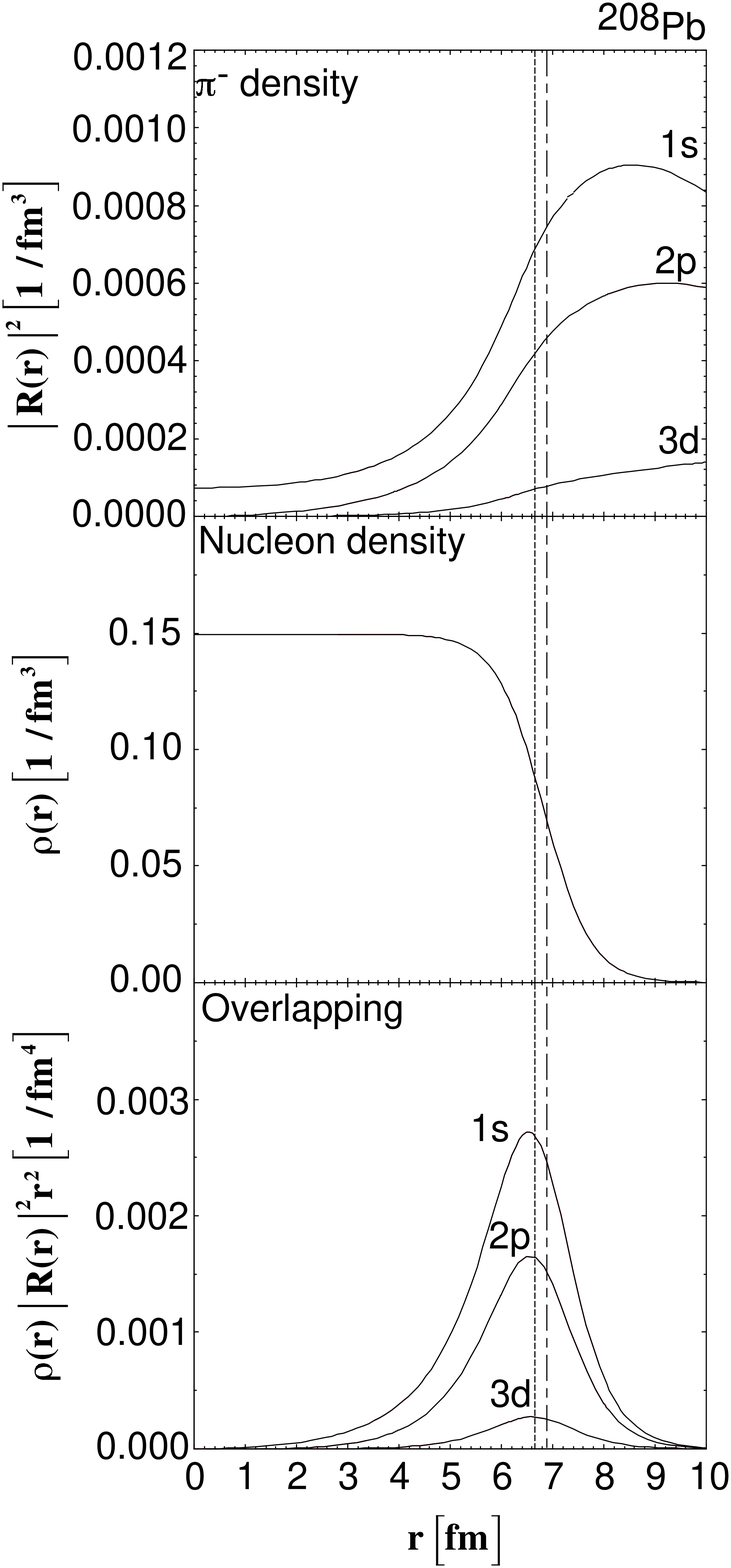,height=14cm}
\vspace{0cm}
\caption{Overlapping probabilities (lower frame) of the $\pi^-$ densities 
(upper frame) with the nucleon densities (middle frame) in typical pionic 
bound states: (Left) $^{16}$O. (Right) $^{208}$Pb. The vertical broken 
lines show the half-density proton radii and the vertical dash-dotted line 
is for the half-density neutron radius in $^{208}$Pb. }
\label{fig:pi-density}
\end{center}
\end{figure}

Figure \ref{fig:pi-density} shows the $\pi^-$ densities ($R_{nl} (r) ^2$), 
the nuclear densities ($\rho (r)$) and the overlapping densities (namely, 
the nuclear densities probed by $\pi^-$), defined as
\begin{equation}
S(r) = \rho (r) \, |R_{nl} (r)|^2  \,r^2,
\end{equation}
for $^{16}$O (1s and 2p) and $^{208}$Pb (1s, 2p and 3d). From these figures 
we notice that the overlapping density is peaked at a radius slightly less 
than the half-density radius, {\it nearly independent of} the nucleus and 
the $\pi^-$ quantum numbers.  This means that the bound $\pi^-$ effectively 
probes a fraction of the full nuclear density ($\rho_0 = \rho (0)$):
\begin{equation}
\rho_e \sim 0.60 \, \rho_0.
\end{equation}
This is a key to intuitively understanding the following results of 
numerical calculations.\\
 
\noindent
{\bf 3. Seki-Masutani relation}\\

Seki and Masutani \cite{SM} emphasized the presence of a strong correlation 
between $b_0$ and Re$B_0$ in describing pionic-atom binding energies and 
widths, which we call the {\it Seki-Masutani relation}. Toki {\it et al.} 
found the same correlation also exists for the deeply bound 1s and 2p 
states in $^{208}$Pb \cite{Toki:88,Toki:89}.  Recently, this correlation 
has been revisited, and the following common relation has been established 
both theoretically and empirically \cite{Geissel:02a}:
\begin{equation}
b_0^* \equiv b_0 + 0.215 \, {\rm Re}B_0 = {\rm constant}.
\label{eq:b0*}
\end{equation}
This means that the binding energies (and widths) are nearly unchanged by 
varying either $b_0$ or Re$B_0$ as long as these parameters are moved 
together to fulfill the above SM relation. In other words, none of $b_0$ 
and Re$B_0$ can be uniquely determined from given binding energies; only 
their combination, $b_0^*$, can be experimentally determined.

\begin{figure}[hbtp]
\begin{center}
\vspace{1cm}
\epsfig{file=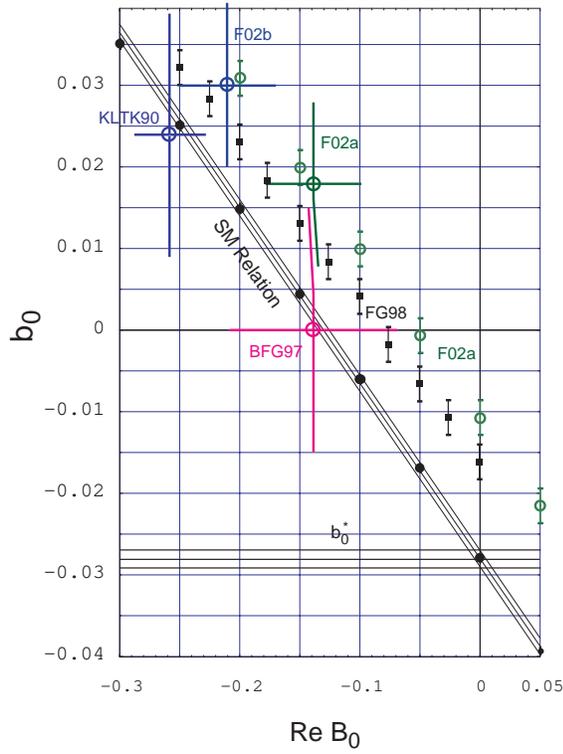,height=10cm}
\vspace{0cm}
\caption{ Seki-Masutani relation between $b_0$ and Re$B_0$. Best-fit values 
of $b_0$ versus Re$B_0$ as a gridding variable, obtained in $\chi^2$ 
minimization using the 1s pionic atom data in 6 $N=Z$ nuclei,  are shown by 
closed circles, whose sizes are equal to the fitting uncertainties. They 
lie on the SM lines:  $b_0^*=b_0 + 0.215 \,{\rm Re}B_0 = -0.0280 \pm 
0.0010$. The \{$b_0$, Re$B_0$\} parameters obtained from global fits of 
KLTK90 \protect\cite{Konijn90}, BFG97 \protect\cite{Batty97}, F02a 
\protect\cite{Friedman:02a} and one of F02b \protect\cite{Friedman:02b} are 
shown by large open circles with both vertical and horizontal error bars. 
Also shown are \{$b_0, {\rm Re}B_0$\} sets from FG98 
\protect\cite{Friedman98} (closed squares) and F02a 
\protect\cite{Friedman:02a} (small open circles), without conversion from 
$b_0$ to $b_0^{\rm F}$ in their convention.}
\label{fig:SM}
\end{center}
\end{figure}

Actually, various global fits gave widely distributed values with large 
errors for these parameters. For instance,
the best-fit values of \{$b_0$,  Re$B_0$\} are \{$0.024(15)$, $-0.26(3)$\} 
in KLTK90 \cite{Konijn90} and \{$0.000(15)$, $-0.14(7)$\} in BFG97 
\cite{Batty97}. With a different definition of $b_0$, namely, $b_0^{\rm F} 
= b_0 - \Delta b_0^{\rm DS}$, in which the double-scattering term ($\Delta 
b_0^{\rm DS}$, to be discussed later) is subtracted, Friedman obtained 
three different sets for \{$b_0^{F}$, Re$B_0$\}: \{$0.018(19)$, 
$-0.14(4)$\} (F02a) \cite{Friedman:02a}, \{$0.030(10)$, $-0.21(4)$\} (F02b) 
\cite{Friedman:02b} and \{$0.020(10)$, $-0.15(4)$\} \cite{Friedman:02b}.
These widely split values are shown by large open circles with both 
vertical and horizontal error bars in Fig.~\ref{fig:SM}. They appear to be 
almost meaningless, contrary to their claim that the parameter Re$B_0$ is 
well determined.
On the other hands, a $\chi^2$ fit with a gridding Re$B_0$ yields a series 
of points, as shown by closed squares (FG98 \cite{Friedman98}) and by small 
open circles (F02a \cite{Friedman:02a}), both of which indicate linear 
relations parallel to the SM correlation.

Recently, the value of $b_0$ with a gridding Re$B_0$ was precisely 
determined by $\chi^2$ fits of the 1s $\pi^-$ binding energies in six light 
symmetric nuclei ($^{12}$C, $^{14}$N, $^{16}$O, $^{20}$Ne, $^{24}$Mg and 
$^{28}$Si) \cite{Geissel:02a}.  The obtained data sets, shown by closed 
circles in Fig.~\ref{fig:SM} (their sizes correspond to the error bars), 
prove the above SM relation (\ref{eq:b0*}) perfectly, yielding a precise 
value of
\begin{equation}
b_0^* = -0.028 \pm 0.001.
\label{eq:b0*value}\end{equation}
It is to be noted that the $b_0^*$ composed of the widely distributed 
values of $b_0$ and Re$B_0$ in each of KLTK90, BFG97, F02a and F02b would 
yield a value, close to $-0.030 \sim -0.034$ (after subtraction of the 
double scattering term in the case of F02a and F02b), which is not so much 
distributed as the individual values of $b_0$ and Re$B_0$. These authors 
seemed to be puzzled by the largeness of their $|{\rm Re}B_0|$ values 
compared with Im$B_0 = 0.055$. On the other hand, they never took seriously 
the other discrepancy  that their $b_0$'s (after subtraction of the double 
scattering term)  take large positive values, meaning a strongly 
attractive $b_0$ in contrast to the nearly vanishing free $\pi$N value 
\cite{Schroeder99,Ericson01}. Such odd consequences from the global fits 
must be artifacts, arising from regarding $b_0$ and Re$B_0$ as independent 
uncorrelated parameters.\\

\noindent
{\bf 4. Seki-Masutani Ansatz and its generalization}\\

Seki and Masutani \cite{SM} showed that the correlation between $b_0$ and 
Re$B_0$ can be understood by imposing an effective replacement of the 
expectation value of the density-quadratic term as,
\begin{equation}
<\rho(r)^2> ~\rightarrow~ \rho_e <\rho(r)>,
\end{equation}
which is expected to hold for any pionic atom state with a common value of 
$\rho_e$ ({\it Seki-Masutani Ansatz}). They claimed an effective nuclear 
density $\rho_e \sim 0.50 \, \rho_0$. We understand that the localization 
of the overlapping densities, $S(r)$, near $r \sim r_c$, as shown in 
Fig.~\ref{fig:pi-density}, is the key to justify the SM Ansatz.

We now examine how the Seki-Masutani Ansatz applies to various  functional 
forms of the potential, $U_s(\rho(r))$.
Let us add the following form to the real part of the s-wave pion-nucleus 
potential:
\begin{equation}
\Delta U_s (r) =  F[\rho(r)] \rho(r),
\label{eq:LDD-potential}
\end{equation}
where $F[\rho(r)]$ is a local-density-dependent (LDD) functional 
coefficient, representing  an additional density-non-linear potential 
involving a local density, $\rho(r)$. We solve the Klein-Gordon equation 
with LDD potentials and compare the numerical results with those without 
invoking LDD assumptions, thereby examining how the effects of the LDD 
potentials are expressed. Following the spirit of SM,  we study how this 
effect is transposed into an effective coefficient,
\begin{equation}
F[\rho(r)] \, \rho(r)  \rightarrow  \overline{F}(\rho_e) \, \rho(r) ,
\label{eq:effective-LDD}
\end{equation}
where $\overline{F}(\rho_e)$ is an effective constant parameter involving 
an effective nuclear density, $\rho_e$. Namely,
\begin{equation}
\overline{F}(\rho_e) = \frac{<F[\rho(r)]\,\rho(r)>}{<\rho(r)>}.
\end{equation}
\\

\noindent
{\bf 5. Polynomial case}\\

Let us first consider the case of a density-polynomial term, which
   can be expressed as
\begin{equation}
{\rm Re}U_s(r) = -(2 \pi/m_{\pi}) \epsilon_1 [b  +  B \rho(r) + D 
\rho(r)^2] \rho(r),
\end{equation}
where
$b = b_0 + (N-Z)/A \times b_1$.
We performed numerical calculations of the 1s binding energies of  $^{16}$O 
and $^{208}$Pb, and found strong correlations between  $B$ and $b$ and 
between $D$ and $b$, as expressed by
\begin{eqnarray}
^{208}{\rm Pb}:&~~&    b + 0.25 B + 0.031 D = ~{\rm const},\\
^{16}{\rm O}:&~~&     b + 0.20 B + 0.025 D = ~{\rm const}.
\end{eqnarray}
The first relation is the well known Seki-Masutani relation for the $B 
\rho(r)^2$ term, and the second relation is a new relation for the $D 
\rho(r)^3$ term.

The effect of the density-polynomial potential is asserted to be equivalent 
to the effect of an effective potential,
\begin{eqnarray}
  U_{SM} (r)&=& -(2 \pi/m_{\pi}) \epsilon_1  \\ \nonumber
  &\times& [b  + 2.826 B \rho_e^{(n=2)}
  + 8.000 D (\rho_e^{(n=3)})^2] \rho (r),
\end{eqnarray}
with $\rho_e$ in units of fm$^{-3}$.
We obtained the following effective densities:
\begin{eqnarray}
\rho_e^{(n=2)} &\sim& 0.088 \sim 0.59 \rho_0,\\
\rho_e^{(n=3)} &\sim& 0.062 \sim 0.41 \rho_0,
\end{eqnarray}
for $^{208}$Pb, and
\begin{eqnarray}
\rho_e^{(n=2)} &\sim& 0.071 \sim 0.42 \rho_0,\\
\rho_e^{(n=3)} &\sim& 0.056 \sim 0.33 \rho_0,
\end{eqnarray}
for $^{16}$O.
The effective density for the $n=2$ term is around $0.5 \, \rho_0$, as 
given by SM, but that for the $n=3$ term is smaller (the pion probes the 
$\rho (r)^3$ component more at a larger $r$).

The effective density does not depend much on the individual states, as we 
have found in the cases of $^{16}$O and $^{208}$Pb. We are therefore 
tempted to expect that the concept of the effective density applies to any 
functional form of $U(\rho(r))$.
\\

\noindent
{\bf 6. Density-dependent $b_1$ parameter}\\

\begin{figure}[hbtp]
\begin{center}
\vspace{1cm}
\epsfig{file=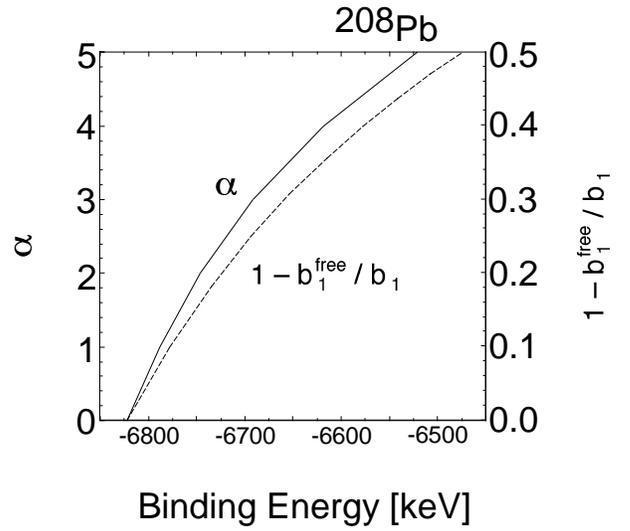,height=7cm}
\vspace{1cm}
\caption{The 1s binding energy calculated in the LDD potential, 
eq.(\ref{eq:LDD-b1}), as a function of $\alpha$, compared with that in a 
conventional potential with $b_1$ as a constant parameter. }
\label{fig:alpha-b1}
\end{center}
\end{figure}

Next, let us consider the case of an isovector s-wave potential which is 
expected to be enhanced due to a partial restoration of the chiral symmetry 
in the nuclear medium \cite{Weise:00,Kienle:01}. We assume the following 
potential according to Weise \cite{Weise:00}:
\begin{equation}
b_1 (r) \Delta \rho(r) = \frac{b_1^{free}}{1 - \alpha \, \rho(r)} \Delta 
\rho(r),
\label{eq:LDD-b1}
\end{equation}
where $\alpha$ is a constant parameter representing the density-dependent 
effect. We examine whether or not the effect of the above LDD potential can 
be replaced by a conventional potential with a constant parameter involving 
an effective density $\rho_e$, as in
\begin{equation}
\overline{b}_1 \Delta \rho(r) =  \frac{b_1^{free}}{1 - \alpha \, \rho_e} 
\Delta \rho(r).
\label{eq:b1-eff}
\end{equation}
Numerical calculations of the binding energies yielded the dependence of 
$B_{1s}$ on the parameter $\alpha$, as shown in Fig.~\ref{fig:alpha-b1}. On 
the other hand, the standard procedure assuming $b_1$ to be a constant gave 
another relation between $B_{1s}$ and $b_1$. The two relations are found to 
be nearly identical, if we take $b_1 = \overline{b}_1$ with
\begin{equation}
\rho_e \sim 0.090 \sim 0.60 \, \rho_0.
\end{equation}
The above finding indicates that the computational results starting from 
the LDD potential invoking a $\rho(r)$-dependent $b_1$ parameter are 
equivalent to the results with a constant-parameter conventional potential; 
the LDD potential can be safely replaced  by a density-linear conventional 
potential with a coefficient, eq.(\ref{eq:b1-eff}).
\\

\noindent
{\bf 7. Double-scattering term}\\

The isoscalar s-wave parameter in the free $\pi$N scattering, $b_0^{\rm 
free}$, was determined experimentally to be nearly zero 
\cite{Schroeder99,Ericson01} in accordance with the Tomozawa-Weinberg 
theorem \cite{Tomozawa,Weinberg}. On the other hand, the double-scattering 
gives a major correction to the isoscalar term, as givenby
\cite{Ericson66,Ericson,Kaiser:01}
\begin{eqnarray}
\Delta b_0^{\rm DS} &=&  - [(b_0^{free})^2 + 2 b_1^2] <\frac{1}{r}>\\
&=&  - [(b_0^{free})^2 + 2 b_1^2] \frac{3}{2 \pi} k_F (\rho),
\label{eq:EE}
\end{eqnarray}
where
$k_F (\rho) = [(3 \pi^2/2) \rho(r)]^{1/3}$.
This correction is local-density dependent, since the correlation function, 
$<1/r>$ ,  is density-dependent and $b_1$ can also be density dependent.

Let us first examine how the above relation changes when we take into 
account the LDD effect on $<1/r>$, while keeping $b_1$ as a constant 
parameter.
We solved the KG equation with
$\Delta b_0^{\rm DS} = - \beta (\rho(r))^{1/3}\, b_1^2$
as an isoscalar part,
and obtained the $B_{1s}$ values in $^{16}$O and  $^{208}$Pb as functions 
of $\beta$, which were compared with  $B_{1s}$ calculated with $\Delta 
\overline{b}_0^{\rm DS}$ as a constant.
Equating the two results in the form
\begin{equation}
\Delta \overline{b}_0^{\rm DS} = - \beta (\rho_e)^{1/3} \, b_1^2,
\end{equation}
we find the effective density, $\rho_e$, is nearly a constant, around $\sim 
0.40 \,\rho_0$ for $^{16}$O and $\sim 0.54 \,\rho_0$ for $^{208}$Pb.
Thus, the coefficient in the double-scattering correction is by a factor of $
(\rho_e/\rho_0)^{1/3} \sim 0.73$ and $\sim 0.81$ (for $^{16}$O and 
$^{208}$Pb, respectively) changed over the conventional formula with 
$\rho_e = \rho_0$, yielding
\begin{equation}
\Delta b_0^{\rm DS} \sim  -1.52 \, b_1^2.
\label{eq:DS-eff}
\end{equation}

Next, we evaluated the full LDD double scattering effect with a 
density-dependent $b_1 (\rho)$. The same form as eq.(\ref{eq:LDD-b1}) with 
$\alpha$ as a running parameter was adopted. The results were compared with 
those calculated with $b_1$ as a constant, as shown in Table~\ref{tab:DS}.
The effective density, $\rho_e/\rho_0 \sim 0.60$, for $^{16}$O and 
$^{208}$Pb agrees well with that found for $b_1 (\rho)$. This means that 
the $\overline{b}_1$ effectively converted from the isovector LDD 
potential term has the same effect as the density-dependent $b_1 (\rho)$ 
in the double-scattering formula, eq.(\ref{eq:DS-eff}).

If we assert that the isoscalar potential strength $b_0^*$, 
eq.(\ref{eq:b0*value}),  is composed of the free $b_0$ (essentially zero), 
the double scattering term and Re$B_0$, we can deduce Re$B_0$ empirically 
as \cite{Geissel:02a}
\begin{equation}
{\rm Re}B_0 = \frac{ b_0^* - \Delta b_0^{\rm DS}}{0.215} \sim -0.038 \pm 0.025.
\label{eq:ReB0}
\end{equation}
This value looks reasonable, when compared with Im$B_0$ = 0.055.
\\

\noindent
{\bf 8. RIA correction}\\

The present procedure can be applied to any kind of additional  potential term.
Friedman \cite{Friedman:02a,Friedman:02b} takes into account the following 
correction
\begin{equation}
\Delta U_{\rm RIA} = -\frac{2 \pi}{m_{\pi}} \frac{3}{5} \frac{m}{M} d_0 
[(1+a \rho)^2 - 1] k_F (\rho)^2 \rho (r),
\end{equation}
where $d_0 = -0.19 ~m_{\pi}^{-3}$ and $a=2.7~ {\rm fm}^3$.  The effect of 
adding $\Delta U_{\rm RIA}$ was calculated and was found to be equivalent 
to invoking an additional isoscalar term
\begin{equation}
\Delta b_0^{\rm RIA} \sim -0.021.
\end{equation}
This correction can be obtained also by setting  $\rho (r) \rightarrow 
\rho_e \sim 0.5\,\rho_0$.
\\

\noindent
{\bf 9. Angle transformation term}\\

Finally, we make some comments on the so called angle transformation term 
as an additional correction to the EE potential,
\begin{eqnarray}
\Delta U_{\rm AT} &=& -(2 \pi/m_{\pi}) [(m_{\pi}/2M) \epsilon_1^{-1} 
\nabla^2[c_0 \rho + c_1 \Delta \rho]\\ \nonumber
&~~~~~+ & (m_{\pi}/M) \epsilon_2^{-1} \nabla^2 (C_0 \rho_n \rho_p)],
\end{eqnarray}
which has been taken into account in the global fits of BFG97, FG98, F02a 
and F02b, but not in the present paper nor in Ref.~\cite{Geissel:02a}. This 
correction behaves like an s-wave potential, though it originates from the 
p-wave parameters. We examined how this correction is expressed in terms of 
the s-wave parameters. We found that the inclusion of this correction 
brings the following changes:
\begin{equation}
\Delta b_0^{\rm AT} \approx +0.003,
\end{equation}
for both $^{16}$O and $^{208}$Pb, and
\begin{eqnarray}
\Delta {\rm Im}B_0^{\rm AT} \approx -0.008~~{\rm for~}^{16}{\rm O},\\
                                             \approx -0.014~~{\rm 
for~}^{208}{\rm Pb}.
\end{eqnarray}
Namely, this correction brings an attractive effect to the real part and a 
narrowing effect to the imaginary part. Thus, the best-fit parameters 
($b_0^*$) with and without this correction have differences, as given 
above. A small difference in $b_0^*$ between the above global fits and the 
present analysis is attributed to this correction. On the other hand, there 
is no sizable effect on $b_1$:  $\Delta b_1^{\rm AT}\approx +0.003$.
\\

\noindent
{\bf 10. Conclusion}\\

We have shown that the Seki-Masutani relation between $b_0$ and Re$B_0$ 
holds for the 1s binding energies of light and heavy pionic atoms, both 
theoretically and empirically. If we admit this fact, the parameter for the 
isoscalar s-wave interaction is well represented by another parameter, 
$b_0^*$, as given in eq.(\ref{eq:b0*},\ref{eq:b0*value}), which takes care 
of the strong correlation between $b_0$ and Re$B_0$. The reason for this 
relation is understood from the fact that the overlapping density of any 
bound $\pi^-$ with the nuclear density is peaked at the radius which is 
slightly smaller than the half-density radius; the nuclear density 
effectively probed by a bound $\pi^-$ is around $0.6 \,\rho_0$.

We have shown that the Seki-Masutani Ansatz can be generalized for any 
functional form of the density-dependent potential:  a LDD term, 
$F[\rho(r)] \, \rho(r)$, can be transposed into a corresponding term 
$F[\rho_e] \, \rho(r)$ with a  constant parameter, where $\rho_e$ is an 
effective parameter. Some important cases are summarized in 
Table~\ref{tab:summary}. In other words, we can say that the conventional 
potential form persists to be valid, because it can represent a LDD 
potential; the constant parameters in the conventional potential are 
regarded as being density-dependent parameters at $\rho \sim \rho_e \sim 
0.60 \, \rho_0$. This also means that the observed enhancement of $b_1$ 
(interpreted as a reduction of the chiral order paramater) in $^{205}$Pb 
\cite{Geissel:02a}, $R = b_1^{free}/b_1 = {f_{\pi}^*}^2/{f_{\pi}}^2 = 
0.78^{+0.13}_{-0.09}$, is for the effective nuclear density of $\rho_e \sim 
0.60 \, \rho_0$, and thus, the reduction that would occur for the full 
nuclear density is expected to be $R \sim 0.63$.    \\

The authors would like to thank Professors R.S. Hayano and P. Kienle for 
the stimulating discussions. This work is supported by the Grant-in-Aid of 
Monbukagakusho of Japan.

\begin{table}[htb]
\begin{center}
\caption{\label{tab:DS}
Relation between the coefficient $\alpha$ representing the 
density-dependent $b_1 (\rho(r))=b_1^{\rm free}/[1-\alpha \rho(r)]$ and the 
equivalent isovector parameter $\overline{b}_1$ in giving the same 1s 
binding energies.
}
\begin{tabular}{lllll}
Nucleus     & $\alpha$ & $1-b_1^{free}/\overline{b}_1$  &$\rho_e$ & 
$\rho_e/\rho_0$\\
\hline
$^{16}$O    &  2    & 0.215  & 0.1075 & 0.63\\
                 &   3    & 0.323  & 0.1077  & 0.63  \\
                  &  4    & 0.428  & 0.1070 & 0.63 \\
\hline
$^{208}$Pb  &  2    & 0.185  & 0.0925 & 0.62\\
                 &   3    & 0.277  & 0.0923  & 0.62  \\
                  &  4    & 0.370  & 0.0925  & 0.62  \\
\end{tabular}
\end{center}
\end{table}

\begin{table}[htb]
\begin{center}
\caption{\label{tab:summary}
Summary of the effective densities, $\rho_e/\rho_0$, obtained for various 
types of local-density-dependent functions, $F[\rho(r)] \rho(r)$.}
\begin{tabular}{llll}
$F[\rho]$ & 1s $^{16}$O & 1s $^{208}$Pb & Remark\\
\hline
$\rho$   & 0.42   & 0.59 & SM relation \\
$1/(1-\alpha \rho)$ & & 0.60 & density-dependent $b_1$ \\
$\rho^{1/3}$ & 0.40 & 0.54 & effective Fermi momentum\\
  $\rho^{1/3}/(1-\alpha \rho)^2$ & 0.63 & 0.62&double-scattering term \\
\end{tabular}
\end{center}
\end{table}

\onecolumn

\twocolumn

\end{document}